\documentstyle[12pt,aasms4]{article}

\begin{document}

\title{Luminous red stars in Local Group dwarf elliptical galaxies: an intermediate-age population?
\altaffilmark{1}}
\author{ David Mart\'\i nez-Delgado and Antonio Aparicio}
\affil{Instituto de Astrof\'\i sica de Canarias, E-38200 La Laguna, Canary Islands, Spain }

\altaffiltext{1} {Based on observations made with the William Herschel Telescope operated on the island of La Palma by the Royal Greenwich Observatory in the Spanish Observatorio del Roque de los Muchachos of the Instituto de Astrof\'\i sica de Canarias.}

\begin{abstract}
In this paper, we show that the optically bright stars above the tip 
of the first red giant branch (TRGB) in the color magnitude (CM) 
diagrams of nearby dwarf elliptical (dE) galaxies, commonly interpreted as indication of the existence of an intermediate-age population, may, in certain circumstances, be an artifact of observational effects on a pure old population system.

For this purpose, model CM diagrams have been computed simulating the observational effects that can be found in different regions of a Local Group dE galaxy. The starting synthetic CM diagram represents a pure old system, with star formation extending only from 15 to 12 Gyr ago. Based on those model diagrams, we analyze the conclusions that a hypothetical ground-based observer might reach concerning the presence of a population of stars above the tip of the red giant branch (TRGB) that could be interpreted as an intermediate-age population, by using two age indicators extensively employed in the literature: i) the bolometric magnitude at the tip of the asymptotic giant branch (AGB) phase, and ii) the ratio of AGB to red giant branch (RGB) stars present in the CM diagram. Our analysis shows that, if observational effects are overlooked, application of both methods would result in finding a fictitious intermediate-age population in the inner regions of the galaxy, where crowding is more severe. 

\end{abstract}

\keywords{Galaxies: elliptical and lenticular; galaxies: stellar content; Local Group; HR diagram; stars: AGB and post-AGB}

\section{INTRODUCTION}
 
Since  Baade (1944) succeeded for the first time in resolving the brightest stars of the dwarf elliptical (dE) 
satellites of Andromeda, the elliptical galaxies remained as simple old stellar systems, with a 
stellar content resembling that of Galactic globular clusters (Population II). Nevertheless, data 
accumulated during the last decades, including the early discovery by Baade himself of blue stars in dEs, provide evidence of varied, and in some cases complex, star 
formation histories among the Local Group dE galaxies (see Hodge 1989 and Freedman 1992a for 
recent reviews). Although there are several indications about the existence of intermediate-age 
population (age between 1 and 10 Gyr) in dEs, the  controversy over the existence of a 
substantial amount of intermediate-age star formation in the Andromeda companions (M32, NGC205, NGC147, NGC185) remains open.

The study of the stellar content of dE galaxies by means of the analysis of the 
distribution of stars in the color-magnitude (CM) diagram provides the most direct way 
of studying their SFH and test whether they have formed stars for 
extended periods of time. Freedman (1989) resolved individual stars in M~32 and found evidence of 
a range of metallicity in the stellar content as well as a significant population of stars more 
luminous than the tip of the RGB (TRGB). Freedman (1992a,b) and Elston \& Silva (1992) discussed 
several possibilities to explain this {\rm optically} bright population (high-metallicity old 
stellar population, old long-period variables, old merged binaries, intermediate-age population), 
reaching the conclusion that these stars probably belong to an intermediate-age AGB population. 
Since then, several studies on the resolved stellar population have found significant numbers of 
these bright stars in the central regions of NGC 205 (Lee 1996), NGC 185 (Lee, Freedman \& Madore 1993; Martinez-Delgado, Aparicio \& Gallart 1996), and NGC~147 (Davidge 1994). All these data suggest that 
the star formation in the Andromeda companions has occurred over an extended period of time and
are also consistent with recent population synthesis results that point to the possibility that elliptical galaxies, dEs in particular, may have a significant intermediate-age stellar population (see Grillmair et al.~1996).
	
Nevertheless,  there is some evidence that the bright stars above the TRGB could be an artifact of the severe crowding in the central regions of these galaxies. Crowding is indeed 
among the most important observational limitations in the study of the stellar populations of these systems. The 
idea that these bright stars are just blends of two or more fainter stars is supported by the 
recent {\it HST} data of M~32 (Grillmair et al. 1996), because the CM diagram of these high-resolution 
observations shows these bright stars much less. Careful analysis of the observational effects in 
NGC~185 (Mart\'\i nez-Delgado et al.~1996) points to similar conclusions.

This paper is devoted to analyzing the observational effects in the upper part of the RGB locus 
and to what extent conclusions about the presence of intermediate-age population in dE galaxies 
are influenced by such effects. Our analysis is based on model CM diagrams\footnote{We use the 
same terms as Gallart et al.~(1996), calling {\it synthetic} CM diagrams those resulting from 
a computer code and {\it model} CM diagrams the same after simulation of observational effects.}. The 
starting point is a 
synthetic diagram simulating a pure old-population system. Crowding effects determined from real 
groud-based observations of a dE galaxy (the data by Mart\'\i nez-Delgado et al. 1996 for NGC~185 
have been used) are simulated in this synthetic diagram. The resulting model diagrams are then 
analyzed using the same methodology widely aplied to real dE galaxies to explore the conclusions that a 
hypothetical ground-based observer would reach concerning the existence of an intermediate-age 
population. Section~2 presents the model CM diagrams. Sections~3 and 4 discuss the effects of an 
intermediate-age population in the model CM diagrams based on the properties of the bright stars above the TAGB. In $\S$5 the results are discussed and a summary of the conclusions is given. 

\section{THE MODEL COLOR-MAGNITUDE DIAGRAMS} \label{cm}

For the present analysis, a synthetic CM diagram with 20000 stars brighter than $M_I=0$ has been 
computed using the Padua stellar evolutionary library (Bertelli et al. 1994 and references therein) following the procedure described in Chiosi, Bertelli \& Bressan (1988) and Gallart et 
al.~(1996). The empirical mass-loss relation for AGB stars by Vassiliadis \& Wood (1993) has been 
used to model the AGB phase. 

The adopted evolutionary scenario corresponds to a hypothetical representation of a dE galaxy in which the stars have formed at a constant rate from 15 to 12 Gyr ago and no more star formation has been produced since then. In other words, the synthetic CM diagram corresponds to a pure old system. The initial mass function (IMF) by Kroupa, Tout \& Gilmore (1993) for the solar neighborhood has been used and a linear chemical enrichment law has been adopted starting in $Z_{i}=0.0001$ at $T_{i}=15$ Gyr and finishing in $Z_{f}=0.004$ at $T_{f}=12$ Gyr (time is expressed in terms of age, i.e., increasing towards the past and 0 being the present time). 

What we want to study is how observational effects affect the quantity of stars located above the TRGB. By observational effects we mean crowding and any other observational limitation distorting or affecting the quality of the CM diagram. Observational effects are mainly of three kinds (Aparicio \& Gallart~1995; Gallart et al.~1996): i) loss of stars; ii) shifts in magnitudes and color indices and iii) increased external photometric errors. A monte-carlo procedure is used to properly simulating all the observational effects, regardless of the nature of the source producing them. We use here the procedure explained in Aparicio \& Gallart (1995) and in Gallart et al. (1996). The reader is referred to those papers for details. Suffice to say here that the method uses the observational effects on a large sample of artificial stars and that it avoids binning or any parametric or analytical modelization. 

A crowding-trial table (see Aparicio \& Gallart 1995) of 72000 stars has been used to simulate the observational effects in the model CM diagrams. Crowding decreases in typical Local Group (LG) dE galaxies as distance to the center increases. To introduce this in our model diagrams, the crowding-trial tables obtained from three regions of NGC 185 (Martinez-Delgado et al.~1996) have been used: i) inner region, $0''\leq a <118''$; ii) intermediate region, $118''\leq a < 236''$ and iii) external region, $236''\leq a < 360''$ where $a$ is the semi-major axis of the ellipses fitting the brightness distribution of NGC~185. It is not the aim of this paper to interpret the stellar content of NGC~185. This galaxy is used only for convenience as a template of the observational effects that can be expected in LG dEs. It is convenient to somehow quantify the observational scenario that drove to the three former crowding-trial tables. Pixel size, seeing and surface brightness can be used as unbiased indication of that sc
enario. Pixel size used was 0.41 arcsec and seeing was ever $\sim0.9$". The average surface brightness of the inner part of NGC~185 is $\sim 21.0$ m/(")$^2$; that of the intermediate region, $\sim 23.5$ m/(")$^2$ and that of the external region, $\sim 25.0$ m/(")$^2$ (Hodge 1963). From now on in this paper and only for convenience, the observational or crowding conditions of the inner part will be termed {\it severe}; those of the intermediate region, {\it moderate} and those of the external region, {\it fair}. Figure 1 shows the initial synthetic CM diagram ({\it a}) and the models obtained after simulating in it the former three levels of crowding. 

\section{THE BOLOMETRIC LUMINOSITY FUNCTION OF AGB STARS}  \label{flb}

The bolometric luminosity funtion of AGB stars has been used as indicative of the presence of an 
intermediate-age stellar population in nearby dE galaxies. The luminosity of the tip of 
the AGB (TAGB) can provide the minimum age of the intermediate-age stars and, therefore, evidence 
for episodes of star formation in the recent past of the galaxy's history. Table~1 shows the bolometric magnitudes  
derived for the TAGB for the Andromeda companion dE galaxies. The references from which the values have been taken are listed in the last column. These values have been considered a sign of the presence of a stellar population as young as just a few Gyr in these galaxies.

Let us focus now in the effect of crowding on the observed cut-off of the AGB bolometric 
luminosity function (LF). To make a quantitative analysis, the bolometric LF has been computed 
for the synthetic CM diagram and the three model CM diagrams shown in Fig. 1 counting stars redder than $(V-I)=1.48$ as proposed by Reid \& Mould (1984). Bolometric corrections were calculated adopting the calibration by Da Costa \& Armandroff (1990). Results are shown in Fig. 2 which displays the bolometric LF of AGB stars for the models of {\it fair}, {\it moderate} and {\it severe} crowding as well as for the synthetic diagram. The latter would correspond to the actual LF, i.e., that which would be found in the absence of any observational effect. 

The LF of the synthetic CM diagram cuts off at $M_{\rm bol}=-4.1$. The interesting point is that the tip of the bolometric LF of model CM diagrams moves to brighter magnitudes when crowding effects are larger, i.e., toward the center of galaxies. For the {\it severe} crowding case, the LF extends at least to $M_{\rm bol}=-4.8$, although a few  
stars are found at still brighter magnitudes ($M_{\rm bol}\sim -5.2$). We adopt the first magnitude because the LF is zero between --4.8 and --5.1. For the {\it fair} and {\it moderate} crowding cases, the TAGB is located at $M_{\rm bol}\sim -4.3$. These results are listed in Table 2: column 1 identifies the model, referring to Fig. 1; column 2 gives the qualitative indication we are using here for the crowding level and column 3 contains the bolometric magnitude of the TAGB derived from the bolometric LF, 

In summary, the simulation shows how observational effects can result in rising the luminosity of the TAGB. If they are not taken into account, a pure old stellar population could be interpreted as being much younger than it actually is. Interestingly, the cut-off of the bolometric LF under {\it severe} crowding conditions is quite similar to those obtained for real galaxies, listed in Table~1.

\section{THE RELATIVE NUMBER OF AGB STARS}  \label{flb}

An alternative quantitative test for the presence of intermediate-age stars derives from 
the observed ratio of the AGB to RGB stars in the CM diagram. The number of stars in these 
evolutionary phases is related to the amount of time that stars spend in them. This time is in 
turn a function of the mass and hence of the age. Therefore, the observed ratio can be compared 
with the prediction of stellar evolutionary theory for different ages to ascertain the presence of an 
intermediate-age population. The ratio of the number of stars one magnitude above and below the TRGB has be 
estimated to be $N_{\rm AGB}/N_{\rm RGB}=0.17$ to 0.25 (Elston \& Silva 1992) when a significant intermediate stellar population is present. The $N_{\rm AGB}/N_{\rm RGB}$ 
ratio determined from observations must be corrected to allow the fact that 20\% of the stars in 
the one-magnitude bin below the TRGB are actually AGBs (Lee 1977; Mould, Kristian \& Da Costa 1983).

The problem arising is that $N_{\rm AGB}/N_{\rm RGB}$ is very sensitive to observational effects. They affect this ratio in three ways, mainly (see Aparicio \& Gallart 1995): i) loss of stars is more important at fainter magnitudes, thus the number of RGB stars could be under-estimated, thereby yielding larger $N_{\rm AGB}/N_{\rm RGB}$; ii) the raising of stars from below to above the TRGB also contribute to larger $N_{\rm AGB}/N_{\rm RGB}$; and iii) shifts of the position of the TRGB strongly affect the former ratio.

To study the aforesaid effects of crowding on conclusions concerning the presence of an 
intermediate-age population in dEs, the $N_{\rm AGB}/N_{\rm RGB}$ ratios have been derived for our 
synthetic and model CM diagrams. First, the position of the TRGB for each diagram has been derived 
from the convolution of the bolometric LF with an edge-detector, a Sobel filter of kernel 
$[-1,0,+1]$ (Madore \& Freedman 1995). The results obtained  for the $N_{\rm AGB}/N_{\rm RGB}$ are given in the last column of Table~2 for the synthetic and model 
diagrams. For the synthetic one, we obtain $N_{\rm AGB}/N_{\rm RGB}=0.08$, in agreement to the fact 
that we are simulating a system without an intermediate-age population. For the model CM 
diagrams, the observed ratios increase as crowding becomes more severe. In the {\it fair} and {\it moderate} crowding cases, $N_{\rm AGB}/N_{\rm RGB}$ are respectively 0.09 and 0.14 and are hence still compatible 
with a system lacking a significant population of intermediate-age stars. But 
$N_{\rm AGB}/N_{\rm RGB}=0.20$ for the {\it severe} crowding case. This result would be interpreted as the galaxy having an important intermediate-age population. But this is again a deceit of observational effects.

\section{CONCLUSIONS}  \label{conc}

Our analysis based on model CM diagrams of a pure old stellar 
population of age $>12$ Gyr shows evidence that the luminous extended giant branch observed in many Local Group dEs could be an artifact of observational effects. We have checked two methods extensively used to test the presence of intermediate-age stars in dEs: the tip of the bolometric LF of AGB stars as an age indicator, and the ratio of the number of stars above and below the TRGB ($N_{\rm AGB}/N_{\rm RGB}$). Application of both methods to our simulated old stellar population results in finding a significant amount of stars resembling an intermediate-age population in regions where crowding is severe, as well as a false gradient of inferred age from the center to the external parts of galaxies. These effects are produced by the raising of the $N_{\rm AGB}/N_{\rm RGB}$ ratio simultaneously with the apparent TAGB moving up to 
brighter luminosities when crowding is severe.

In the light of these results, the simple presence of bright stars populating the region of the CM 
diagram above the TRGB of nearby galaxies should not be considered as unambiguous proof of the existence of an intermediate-age population in these systems if observational effects are not under control. Furthermore, gradients in the number of these bright stars as a function of the galactocentric distance could also be the result of changing crowding conditions and care should be 
taken in interpreting the former as a gradient in the age of the stellar population. Careful analysis of the observational effects in each particular case is called for to test whether or not an intermediate-age population is actually present in the galaxy and what its distribution is. Unfortunately, observational effects are of random nature. They depend on many subtle factors difficult to take under control and the simple knowledge of the surface brightness, the seeing or the pixel sampling is probably not enough to characterize them. The only secure procedure is, in our oppinion, performing monte-carlo simulations using as a reference a crowding-trial table generated in the same frames under analysis and containing information for a large number of artificial stars in order of minimizing undesirable systematic effects on the simulations (see Aparicio \& Gallart 1995)

The above reasoning shows that severe crowding is a sufficient condition to find a large number of bright stars over the TRGB. But it is still possible that dEs do have a significant intermediate-age population, and that some of the 
bright stars over the TRGB are actually intermediate-age stars. However, indications exist 
that the intermediate-age population might best be looked for in the AGB red-tails appearing in many CM diagrams of dEs. 

\acknowledgments This work has been substantially improved after fruitful discussions with Dr. Cesare Chiosi, Dr. Gary Da Costa, Dr. Carme Gallart and Dr. Barry Madore, to whom we are grateful. This work has been financially supported by the Instituto de Astrof\'\i sica de Canarias (grant P3/94) and the Direcci\'on General de Investigaci\'on Cient\'\i fica y T\'ecnica of the Kingdom of Spain (grant PB94-0433).

\newpage

\figcaption[figure1.ps]{Synthetic ({\it a}) and model ({\it b} to {\it d}) CM diagrams for a pure old system with star formation from 15 to 12 Gyr ago. Model diagrams are obtained simulating in the synthetic diagram the observational effects found in a typical LG dE galaxy. {\it None}, {\it fair}, {\it moderate} and {\it severe} reffer to these observational effects (see the text for a quantitative definition of these terms). \label{crowding}}

\figcaption[figure2.ps] {Bolometric luminosity functions of bright, red stars obtained for the synthethic CM diagram and for the three model CM diagrams plotted in Fig. 1. Full line and full dots correspond to the LF of the synthetic diagram. Dashed line and open dots correpond to the LF of each model diagram, after the observational effects have been simulated. {\it Severe}, {\it moderate} and {\it fair} reffer to the observational effects (see the text for a quantitative definition of these terms). \label{fbol}}


\begin{references}
\reference{apar95} Aparicio, A., \& Gallart, C. 1995, \aj, 110, 2105
\reference{baade} Baade, W. 1944, \apj, 100, 147
\reference{bertelli} Bertelli, G., Bressan, A., Chiosi, C., Fagotto, F., \& Nasi, E. 1994, A \& AS, 106, 271
\reference{chiosi} Chiosi, C., Bertelli, G., \& Bressan, A. 1988, \aap, 196, 84
\reference{dacosta2} Da Costa, G. S., \& Armandroff, T. E. 1990, \aj, 100, 162
\reference{davidge} Davidge, T. 1994, in  ESO conference and Workshop Proc. 51,
The Local Group: Comparative and Global Properties, ed. A. Layden, R.C. Smith \& J. Storm (Garching:ESO), 92
\reference{elston} Elston, R., \& Silva, D. 1992, \aj, 104, 1360
\reference{freedman} Freedman, W. L. 1989, \aj, 98, 1285
\reference{freedman92a} Freedman, W. L. 1992a, in  The stellar population of Galaxies, ed. B. Barbuy \& A. Renzini (Dordrecht: Kluwer), 297
\reference{freedman92b} Freedman, W. L. 1992b, \aj, 104, 1349
\reference{gallart} Gallart, C., Aparicio, A., Bertelli, G., \& Chiosi, C. 1996, \aj, 112, 1950
\reference{grillmair} Grillmair, C. J., Lauer, T. R., Worthey, G., Faber, S.M., Freedman, W. L., Madore, B. F., Ajhar, E. A., \& Baum, W. A. 1996, \aj, 112, 1975 
\reference{hodge2} Hodge, P. W. 1963, \aj, 68, 691
\reference{hodge} Hodge, P. W. 1989, \araa, 27, 139
\reference{kroupa} Kroupa, P., Tout, C. A., \& Gilmore, G. 1993, \mnras, 262, 545
\reference{lee} Lee, S.W. 1977, \aap S, 27, 381
\reference{leemg} Lee, M. G., Freedman, W. L., \& Madore, B. F. 1993, \aj, 106, 964
\reference{lee96} Lee, M. G. 1996, \aj, 112, 1438
\reference{madore} Madore, B. F., \& Freedman, W. L. 1995, \aj, 109, 1645
\reference{martinez} Mart\'\i nez-Delgado, D., Aparicio, A., \& Gallart, C. 1996, in preparation
\reference{mould} Mould, J. R., Kristian, J., \& Da Costa, G. S. 1983, \apj, 270, 471
\reference{reid} Reid, N., \& Mould, J. 1984, \apj, 284, 98
\reference{vw} Vassiliadis, E., \& Wood, P.R. 1993, \apj, 413, 641


\end{references}
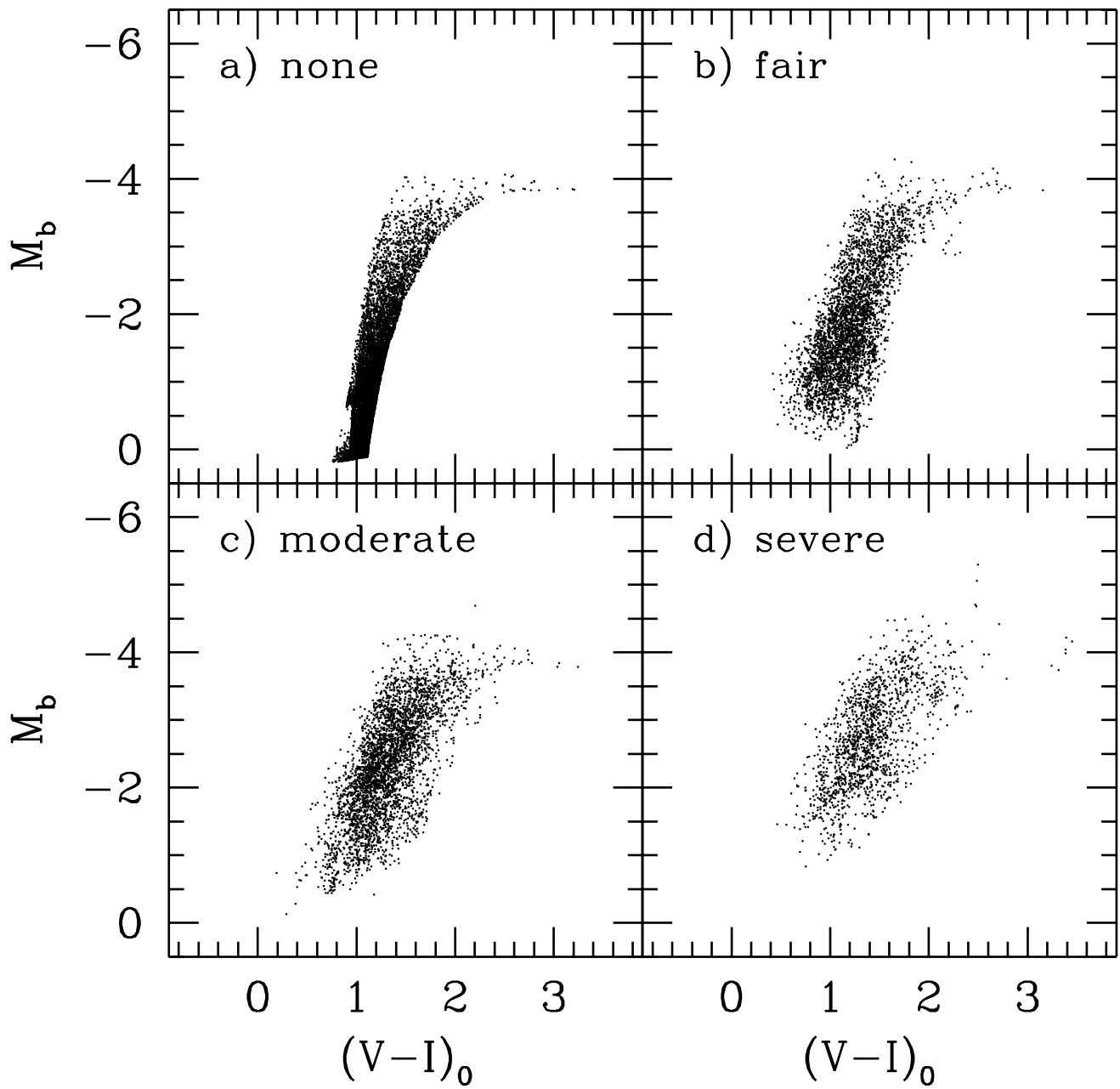
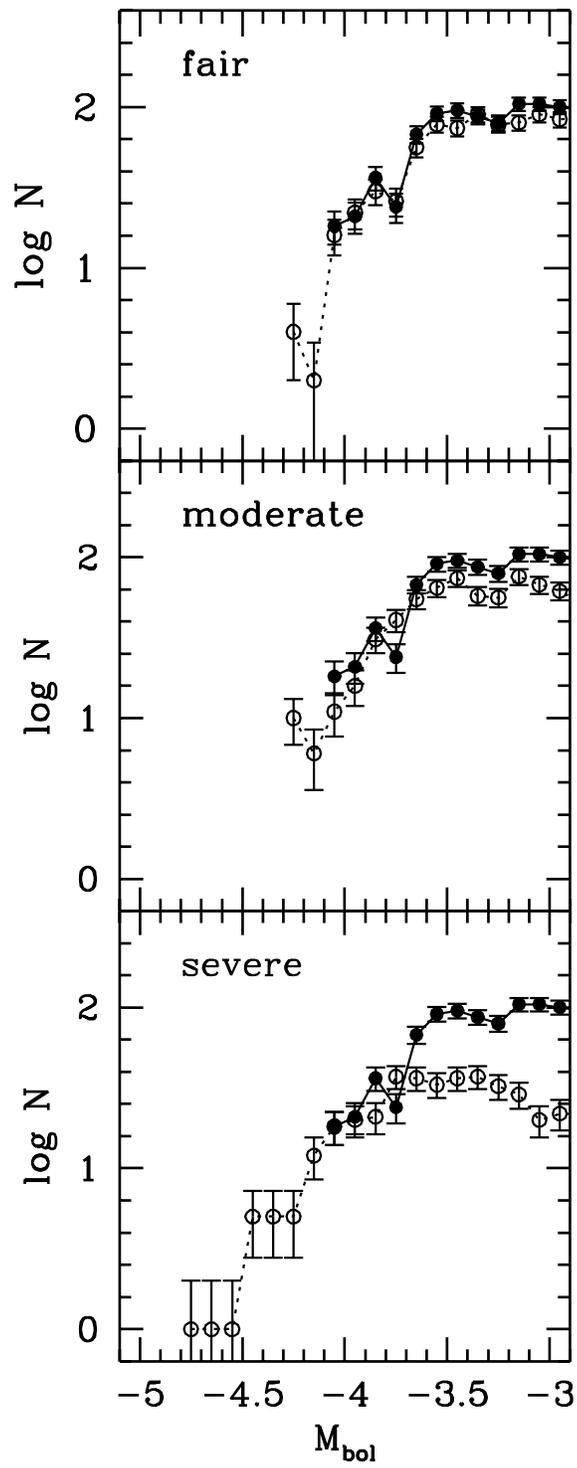
\end{document}